\documentstyle[ fleqn,epsfig]{article}
\newcommand{\AmS}{{\protect\the\textfont2
    A\kern-.1667em\lower.5ex\hbox{M}\kern-.125emS}}
\begin{document}
\title{
\rightline{\small  UL--NTZ 08/97} 
\vspace{-10pt}
\rightline{\small  HUB--EP--97/14} 
Measuring the interface tension when the electroweak phase transition 
becomes weak}
\author{M.~G\"urtler$^1$\thanks{guertler@tph204.physik.uni-leipzig.de}, 
  E.-M.~Ilgenfritz$^2$\thanks{ilgenfri@pha1.physik.hu-berlin.de}
  and
  A.~Schiller$^1$\thanks{schiller@tph204.physik.uni-leipzig.de} \\
  {\it $^1$ Institut f\"ur Theoretische Physik, Universit\"at Leipzig, Germany} \\
  {\it $^2$ Institut f\"ur Physik, Humboldt-Universit\"at zu Berlin, Germany}
  }
\date{April 16, 1997}
\maketitle
\begin{abstract}
  We measure the interface tension near the phase transition endpoint of the
  $3d$ $SU(2)$--Higgs model. The tunnel correlation length method is used and
  compared to other approaches. A modified scaling behaviour for the mass gap as
  function of the transverse area is proposed.  \vspace*{1cm}
\end{abstract}

\section{Introduction}

It is now almost established that the symmetry restoring electroweak phase
transition changes into a smooth crossover if the mass of the so far elusive
Higgs particle would be equal to the W-mass (or less within 10 per cent)
\cite{kajantieprl96,karsch96,aoki,preparation}.  This emerges from lattice
studies in the effective 3-dimensional model as well as from 4-dimensional
simulations, the latter so far only with a rough lattice spacing of order
$O(1/(2T))$ with temperature $T$.

In order to quantify the strength of the first order transition near to the
critical Higgs mass several quantities have been considered.  Measuring the
discontinuity of the Higgs condensate is practically tantamount (in three
dimensions) to calculating the latent heat.  Both quantities are easy to obtain
but are very sensitive to finite volume effects. It is even more difficult to
measure the interface tension for weak transitions.  This we have observed in
our recent work \cite{wirNP97} where we have estimated the interface tension for
a Higgs mass near to its critical value.

In the literature mainly three methods are discussed to extract the interface
tension from Monte Carlo studies. In the present work we shall employ and
critically discuss the tunneling correlation length method
\cite{privman,muenster}, which has been used so far only in the analysis of
$4$--dimensional simulations \cite{bunk94,csikorpl95} in the context of the
electroweak phase transition.

In section 2 we define the model and discuss the methods used to extract the
interface tension. The results obtained by the tunneling correlation method are
presented and critically discussed in section 3. They are put into perspective
with all available lattice measurements for arbitrary Higgs masses in section 4,
which contains also our conclusion on the capability of the method.

\section{The Model and How to Measure the Interface Tension}

The model under study is a $3$--dimensional $SU(2)$--Higgs system with one
complex Higgs doublet of variable modulus. The gauge fields are represented by
unitary $2 \times 2$ link matrices $U_{x,\alpha}$ and the Higgs fields are
written as $\Phi_x = \rho_x V_x$. $\rho_x^2= \frac12{\mathrm tr}(\Phi_x^+\Phi_x)$ is the
Higgs modulus squared, $V_x$ an element of the group $SU(2)$, $U_p$ denotes the
$SU(2)$ plaquette matrix. The lattice action is
\begin{eqnarray}
  S  &=& \beta_G \sum_p \big(1 - {1 \over 2}{\mathrm tr} U_p \big) -   
  \beta_H \sum_{l=(x,\alpha)}
  {1\over 2}{\mathrm tr} (\Phi_x^+ U_{x, \alpha} \Phi_{x + \alpha})
  \nonumber \\
  &  & +  \sum_x  \big( \rho_x^2 + \beta_R (\rho_x^2-1)^2 \big)
  \label{eq:latt_action}
\end{eqnarray}
(summed over plaquettes $p$, links $l$ and sites $x$), with the gauge coupling
$\beta_G$, the lattice Higgs self--coupling $\beta_R$ and the hopping parameter
$\beta_H$. To relate the lattice couplings to continuum quantities the notations
of \cite{wirNP97} are used.
 
Let us recall the procedures used for extracting the interface tension of the
electroweak phase transition.  The first one, the two--coupling method, requires
a careful extrapolation to the limits $\beta_{Hc}\pm \varepsilon$ of simulations
in two coupled subvolumes \cite{csikorpl95}--\cite{hein} supposed to be kept in
the two phases.  The variants differ in the way this limit is taken.  In our
version of the method \cite{emias94} we have used a simultaneous multihistogram
technique for a system consisting of two parts in order to estimate the free
energy difference between the homogeneous and mixed states at $\beta_{Hc}$.
 
The majority of results for the interface tension has been obtained with the
second method based on the two--state signal in the histogram of an order
parameter like quantity $o$ \cite{csikor,KajantieNP96,wirNP97,aoki}. Here the
whole system is simulated at the pseudocritical point. The bulk variable under
consideration is, say, the volume average of the modulus squared $o=
\overline{\frac12{\mathrm tr}(\Phi^+ \Phi)}$ or the average link
$o=\overline{\frac12 {\mathrm tr}(\Phi^+ U \Phi)}$.  Usually, the interface
tension $\alpha$ is estimated comparing the minimum and the two maxima of the
doubly peaked histogram $p(o)$ by
\begin{equation}
  \frac{\alpha a^2 } {T_c}= \frac{1} {2 L^2} \log \frac{p_{\mathrm max}}
  {p_{\mathrm min}}
  \  ,
  \label{eq:one}
\end{equation}
where $L$ is a typical linear extension of a surface eventually dividing the
system into different phases (the smaller extensions of a cylindrical system),
$a$ denotes the lattice spacing and $T_c$ the corresponding transition
temperature. Different entropy factors in the thermodynamic weight of the mixed
state have been ignored for simplicity in (\ref{eq:one}), but are necessary to
extract a reasonable estimate for the interface tension, in particular if
histogram data of lattices with various geometries are used simultaneously
\cite{iwasaki}.  Applying a formula like (\ref{eq:one}) one tacitly assumes that
({\it i}) the order parameter $o$ is well--chosen in order to yield a histogram
with clearly separated maxima and a broad minimum in between, ({\it ii}) the
case of equal height of the maxima is near to the phase equilibrium, ({\it iii})
a minimal surface spanning through the lattice separates the pure phases from
each other.

Lattice studies of the electroweak transition in the case of the Standard Model
for realistic values of the Higgs mass are known to possess a transition which
is very asymmetric and weak.  Therefore the conditions ({\it i}) to ({\it iii})
are hardly fulfilled.  By ''asymmetric'' we mean that the fluctuations of
suitable order parameters $o$ are stronger in one (the Higgs) phase than in the
other (symmetric) phase.  This fact makes it more and more difficult to use
histogram methods based on (\ref{eq:one}) when approaching the critical Higgs
mass.

In order to cope with overlapping histograms we have recently proposed a method
to split histograms $p(o)$ into pure phase and mixed phase contributions
\cite{wirNP97} even under realistic circumstances of a very weak transition.
This has enabled us to employ the histogram reweighting technique to find the
pseudocritical coupling by the equal weight criterion. As a by--product, we have
also obtained the latent heat from the variation of both pure phase
thermodynamic weights near to the transition point and the thermodynamic weight
of mixed states.  Extracting, however, the interface tension from the relative
weights of pure and mixed phases still depends on the simplifying assumption
({\it iii}) above.  Collecting data from lattices with different aspect ratios
and extrapolating in the smallest linear extension to infinity we \cite{wirNP97}
obtained at the physical Higgs mass $m_H= 64.77$~GeV (which corresponds to
$T_c=150.9$~GeV) for the $SU(2)$--Higgs theory without fermions an estimate of
$\alpha/T_c^3 = 2.1 \times 10^{-4} $.

All methods discussed so far are focusing on changes of (volume or subvolume)
averaged variables $o$ in mixed phase systems with minimal interfaces.  There is
a third method \cite{privman,muenster} to deduce the interface tension $\alpha$
from a tunneling correlation length $\xi_{\mathrm tunnel}$ at the phase
transition.  To be more precise, it is the dependence of this correlation length
on the geometry of the system which allows to extract $\alpha$.  A first test of
the credibility of the method in the case of the electroweak phase transition
has been undertaken in Ref.~\cite{bunk94} far from the critical Higgs mass and
in Ref.~\cite{csikorpl95} at even smaller Higgs mass, both within the
$4$--dimensional framework.
 
In contrast to measurements of the temperature dependent Higgs mass on both
sides of the transition \cite{wirNP97} (which requires to separate pure phase
samples) the emphasis is here on correlations due to different phases in
coexistence.  The tunneling correlation length is measured in a very elongated
volume $L^2 \times L_z$, stretched along the $z$--direction.  In
$4$--dimensional simulations the additional Euclidean temporal extent is
understood to represent the (inverse) temperature and cannot be modified in
practice.  The connected correlator $C_{\mathrm conn}(z_1-z_2)$ of $o(z)=\sum_x
\delta_{x_3,z} ~{\mathrm tr}~(\Phi_x^+ \Phi_x)$ between two equal--$z$ slices defines the
correlation length $\xi_{\mathrm tunnel}$ which is expected to vary with the
transversal extent of the lattice as
\begin{equation}
  \xi_{\mathrm tunnel} \propto 
  \exp\left( \alpha_3 A  \right), \ \ \ \ \ \ A=(a L)^2 \ \  . 
  \label{eq:two}
\end{equation}
The interface tension of the original 4--dimensional theory $\alpha$ is related
to $\alpha_3$ through $ {\alpha}= \alpha_3 T_c$.
 
Using a semiclassical expansion for a scalar $\varphi^4$ theory in $3d$
including qua\-dra\-tic fluctuations around a kink solution \cite{muenster}, the
mass (energy) gap $m = \xi_{\mathrm tunnel}^{-1}$ has been calculated (in
lattice units) as
\begin{equation}
  m_{\mathrm lat}=m~a = C ~\sqrt{\frac{\alpha a^2}{T_c}}  
  ~\exp \left( - \frac{\alpha L^2 a^2 }{T_c} \right) 
  \label{eq:three}
\end{equation}
with no additional $L$ dependence in front of the exponent.  Therefore,
expressed in terms of the dimensionless parameter
\begin{equation}
  x=\frac{\alpha L^2 a^2 }{T_c} \ \  ,
  \label{eq:three1}
\end{equation}
the following scaling behaviour is expected to hold for the tunneling
correlation lengths (if they are measured in units given by the transverse
lattice size)
\begin{equation}
  m_{\mathrm lat} ~L = C \sqrt {x} ~\exp (- x) \ \  .
  \label{eq:scaling}
\end{equation}

Strictly speaking, this result is valid for systems within the universality
class of the Ising model, but it has been confronted with the $4$--dimensional
$SU(2)$--Higgs model in Ref.~\cite{bunk94}.  In this first application in the
electroweak context (at smaller Higgs mass and, consequently, higher $\alpha$)
it has been shown that the simple perturbative one--loop result (\ref{eq:three})
is reached from above with increasing transverse lattice size and is valid only
beyond $x \simeq 1$ (cf. Fig.~\ref{fig:four} below).  It has been argued already
in \cite{brezin} that higher order corrections may lead to a pre--exponential
power in $L$ with an exponent different from zero in (\ref{eq:three})
\footnote{This has been confirmed in a recent two--loop calculation
  \cite{private_muenster}.}.  Concentrating on the {\it roughening} of the
interface in a capillary wave model beyond the Gaussian approximation
\cite{caselle} it has been found that (\ref{eq:three}) gets a correction factor
$(1 + O\left(T_c/(\alpha L^2 a^2)\right))$.

Recall also that in all derivations it has been assumed that the tunneling
correlation length is much larger than the typical correlation lengths in the
pure phases.  In our recent studies at a physical Higgs mass of roughly $65$~GeV
\cite{wirNP97} we have measured the Higgs correlation lengths near to the phase
transition in the pure phases as $\xi_{\mathrm broken}/a=13.40(41)$ and
$\xi_{\mathrm symm}/a =9.71(29)$ which are not so small compared to the
tunneling correlation length $1/(m_{\mathrm lat} a)$ as one will see later.
Furthermore we have visualised there a typical mixed--phase configuration which
had very rough interfaces separating different phases.

Therefore, in order to extract an interface tension at this very weak first
order transition (with very rough interfaces) we assume that the tunneling mass
gap can be parametrised in a more general form as function of the transverse
extent $L$
\begin{equation}
  m_{\mathrm lat} ~L = c ~L^{\gamma} ~\exp \left(- ~\alpha_{\mathrm lat} ~L^2
  \right) \ \  ,
  \label{eq:four}
\end{equation}
with the fit parameters $c$, $\gamma$ and $\alpha_{\mathrm lat}$.  Using the
$3d$ continuum gauge coupling $g_3^2=4/(\beta_G a)$, we can put $\alpha_{\mathrm
  lat}$ into relation to the $3d$ dimensionless interface tension
$\alpha_3/g_3^4$ by
\begin{equation}
  \label{eq:alpha_lat}
\frac {\alpha_3} {g_3^4} 
=\left(\frac{\beta_G}4\right)^2  \alpha_{\mathrm lat}  \ \  .
\end{equation}
As a check of this assumption we have to compare the interface tension with the
result of other methods when these are available.

\section{Results and Discussion}

The lattice model (\ref{eq:one}) is used as in \cite{wirNP97}, in particular
with the same update algorithm as described there.  We are dealing with the two
cases corresponding to the Higgs masses $M_H^*=70$ and $M_H^*=57.4423$~GeV
(denoted in the following by 57~GeV). These cases correspond to
${\lambda_3}/{g_3^2} \approx 0.095703$ and ${\lambda_3}/{g_3^2} \approx
0.0644457$, respectively, ($\lambda_3$ is the $3d$ continuum Higgs self coupling)
via
\begin{equation}
  \frac{{\lambda_3}}{{g_3^2}}=\frac18 \Big(\frac{M_H^*}{80\;
  \mbox{GeV}}\Big)^2 \ .
  \label{MH*}
\end{equation}
The Higgs mass $M_H^*$ (in GeV) differs numerically only slightly from the
physical Higgs mass $m_H$ in the $4d$ theory without top.

The smaller Higgs mass is chosen in accordance with the work \cite{KajantieNP96}
(in their notation referred to as $m_{H}^*=60$~GeV).  Correlation function
measurements have beeN Taken after each 10th iteration. The maximum of the
integrated autocorrelation time for this quantity was about $26$ at the smaller
Higgs mass and $16$ at the larger Higgs mass (each in case of the largest
measured transverse size).

To obtain the tunneling correlation length $\xi_{\mathrm tunnel}$ requires some
tuning. At first, the appropriate hopping parameter value $\beta_H$ has to be
tuned, separately for each transverse size of the system while keeping near to
the bulk critical value $\beta_{Hc}$, to the maximum of the tunneling
correlation length.  The actual longitudinal size of the lattice has been chosen
three to four times larger than the correlation length one is going to measure.

In the tables we quote the statistics for all lattice geometries $L^2 \times
L_z$ at the respective $\beta_H$ (having the maximal tunneling correlation
length) and its corresponding value $m_{\mathrm lat}$. We indicate also the
inverse transverse correlation length $m_{\mathrm \perp\ lat}$ and the total
statistics which went into the search for the minimum.  Additionally, we have
checked at the larger Higgs mass for geometries $12^2 \times 96$ and $20^2
\times 128$ that $m_{\mathrm lat}$ does not change within the errors for
larger $L_z$.
\begin{table}[!thb]
  \centering
  \begin{tabular*}{120mm}{@{\extracolsep{\fill}}|c|c|c|c|c|c|}
    \hline
    $ L^2 \times L_z$    &
    $ m_{\mathrm lat} $  &
    $\#$ msmts         &
    $\beta_H$           &
    $ m_{\perp \mathrm lat} $ &
    total $\#$ msmts  \\ \hline
    $4^2  \times 32 $ & .2180(106)  &  8000 & .343600 &  ---      & 36000 \\
    $6^2  \times 64 $ & .1543(117)  & 10000 & .343000 &  ---      & 24000 \\
    $8^2  \times 64 $ & .1165(081)  &  6000 & .342700 &  ---      &  8000 \\
    $10^2 \times 64 $ & .09150(531) &  6000 & .342700 & .1008(29) & 10000 \\
    $12^2 \times 64 $ & .07060(339) &  6000 & .342694 & .0782(21) & 12000 \\
    $14^2 \times 96 $ & .05200(380) & 10000 & .342700 & .0632(21) & 19000 \\
    $16^2 \times 128$ & .04195(286) & 10000 & .342688 & .0509(15) & 21000 \\
    $18^2 \times 128$ & .03517(331) & 12000 & .342686 & .0425(14) & 20000 \\ 
    \hline
  \end{tabular*}
  \caption{Statistics at $M_H^*=57$~GeV}
  \label{tab:57}
\end{table}
\begin{table}[!thb]
  \centering
  \begin{tabular*}{120mm}{@{\extracolsep{\fill}}|c|c|c|c|c|c|}
    \hline
    $ L^2 \times L_z$    &
    $ m_{\mathrm lat} $  &
    $\#$ msmts         &
    $\beta_H$           &
    $ m_{\perp \mathrm lat} $ &
    total $\#$ msmts  \\ \hline
    $4^2  \times 128 $ & .276(6)     &  3000 & .345000 &  ---      & 18400 \\
    $6^2  \times  64 $ & .1933(044   & 10000 & .344200 &  ---      & 60000 \\
    $8^2  \times 64  $ & .1489(053)  & 20000 & .343800 &   ---     & 40000 \\
    $10^2 \times 64  $ & .1174(043)  & 25000 & .343540 & .1345(17) & 80000 \\
    $12^2 \times 64  $ & .09530(278) & 50000 & .343540 & .1080(13) &110000 \\
    $14^2 \times 64  $ & .08031(267) & 25000 & .343540 & .0889(16) & 25000 \\
    $16^2 \times 64  $ & .06910(154) & 25000 & .343540 & .0761(09) & 40000 \\
    $20^2 \times 64  $ & .05367(163) & 15000 & .343560 & .0599(12) & 51000 \\ 
    \hline
  \end{tabular*}
  \caption{Statistics at $M_H^*=70$~GeV}
  \label{tab:70}
\end{table}

\begin{figure}[!htb]
  \centering
  \epsfig{file=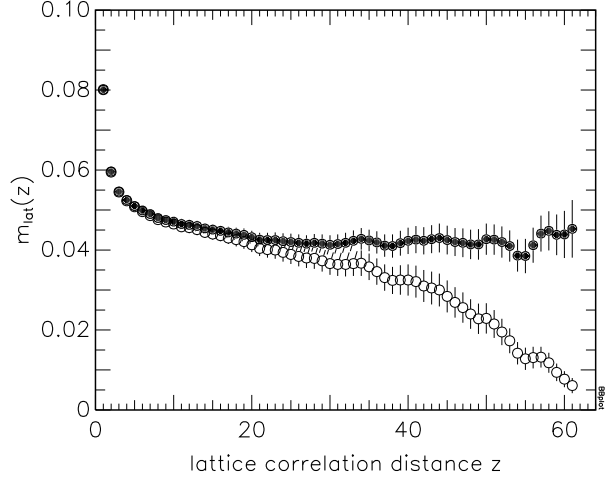,width=8.0cm,angle=0}
  \caption{\sl Example of the local lattice mass at $M_H^*=57$~GeV 
    and $16^2 \times 128$ as function of z}
  \label{fig:one}
\end{figure}

\begin{figure}[!htb]
  \centering
  \epsfig{file=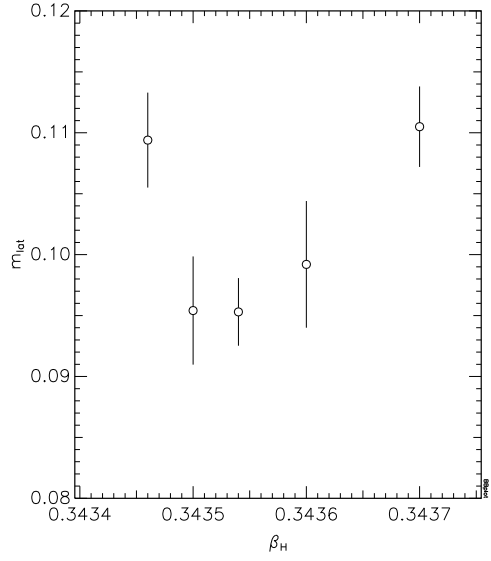,width=6.5cm,angle=0}
  \caption{\sl Inverse correlation length at $M_H^*=70$~GeV 
    and $12^2 \times 64$ as function of  $\beta_H$}
  \label{fig:two}
\end{figure}

In order to extract the correlation length $\xi_{\mathrm tunnel}$ we first check
by inspection whether the local mass has a plateau. We define a local lattice
mass $m_{\mathrm lat}(z)$ at correlation distance $z$ through a fit of three
subsequent values $C_{\mathrm conn}(z-1)$, $C_{\mathrm conn}(z)$ and $C_{\mathrm
  conn}(z+1)$ of the correlation functions of $o(z)=\sum_x \delta_{x_3,z}
~{\mathrm tr}~(\Phi_x^+ \Phi_x)$ to a hyperbolic cosine shape $A \ (\exp(- m_{\mathrm
  lat}(z) \ z) + \exp( -m_{\mathrm lat}(z) \ (L_z - z) )) $.  This ansatz is
essential in order to observe a plateau in the local mass versus $z$.  In
Fig.~\ref{fig:one} $m_{\mathrm lat}(z)$ is shown for one particular example.  We contrast this
with a local mass defined through a single exponential fit to the three
neighbouring values of $C_{\mathrm conn}(z)$ above.  No plateau at all can be
identified using the latter definition of a local mass.  Finally, the (inverse
of the) correlation length $\xi_{\mathrm tunnel}$ is obtained by a global
hyperbolic cosine fit over the plateau range that we have identified.
The behaviour of the inverse correlation length near to its lowest value is
demonstrated in Fig.~\ref{fig:two}.

\begin{figure}[!htb]
  \centering
  \epsfig{file=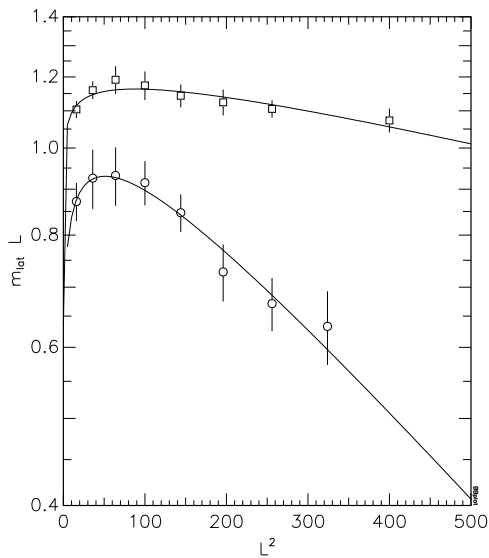,width=6.5cm,angle=0}
  \caption{\sl Fits for the inverse tunneling correlation lengths at $M_H^*=57$
   ~GeV (circles) and $M_H^*=70$~GeV (squares)}
  \label{fig:three}
\end{figure}

In Fig.~\ref{fig:three} we show the inverse tunneling correlation length
multiplied by the transverse extension, $m_{\mathrm lat}\ L$ as function of the
transverse lattice area $L^2$ for both values of the Higgs mass under
discussion.  The strongly different exponential slopes reflect the weakening of
the phase transition.  The curves correspond to a least square fit with the
ansatz (\ref{eq:four}).

From this fit we extract the dimensionless $3d$ interface tensions
$\alpha_3/g_3^4=0.0224(56)$ for the lower Higgs mass $M_H^*=57$~GeV, and
$\alpha_3/g_3^4=0.0049(18)$ for $M_H^*=70$~GeV.  The first number is in very
good agreement with the value presented in Fig.~15 of \cite{KajantieNP96} (and
-- with the use of eqs. (11.6) and (2.8) there -- translated into)
$\alpha_3/g_3^4=0.0217(22)$.

Using these numbers we find the interface tensions $\alpha$
\begin{equation}
  \frac{\alpha}{T_c^3}=  3.24(80) \times 10^{-3} , \ \ \ M_H^*=57 {\mathrm{GeV}}
  \label{eq:alpha57}
\end{equation}
and  
\begin{equation}
  \frac{\alpha}{T_c^3}=  70(26) \times 10^{-5}, \ \ \ M_H^*=70 {\mathrm{GeV}}
  \label{eq:alpha70}
\end{equation}
corresponding to the $4$--dimensional model without top quark.  Taking into
account the top quark the numbers change to $3.57(89)\times 10^{-3}$ and
$77(29)\times 10^{-5}$.  The relations between $3$--dimensional and
$4$--dimensional quantities have been obtained as in \cite{wirNP97} using (in
the same notations as there) table~\ref{tab:correct} which is recalculated
here for $M_H^*=57$~GeV.
\begin{table}[!thb]
  \centering
  \begin{tabular*}{100mm}{@{\extracolsep{\fill}}|c|r|r|}
    \hline
                      &  $M_H^*=57$~GeV     &  $M_H^*=57$~GeV   \\  
                      &  without fermions   & with top          \\ \hline
    $\Delta_g$        &  -0.01322           &  $6.712 \times  10^{-5}$  \\
    $\Delta_\lambda$  &  -0.02383           &  1.494            \\
    $\Delta_\nu$      &  -0.02938           &  0.9443            \\   
    $\Delta_Y$        &                     & -0.01301           \\   \hline
  \end{tabular*}
  \caption{Corrections in eq.~(6) of [5]}  
  \label{tab:correct}
\end{table}
Though the top contribution apparently changes the interface tension only
insignificantly ($\Delta_g$ is small) some of the fermionic one-loop corrections
to the $4d$ couplings are already too large and the other physical numbers
should be taken with great care.  The reported pseudocritical $\beta_H$ values
for the largest transverse sizes from tables~\ref{tab:57} and \ref{tab:70} are
related to critical temperatures $T_c=127.2$ and $150.1$~GeV and physical Higgs
masses $m_H=52.35$ and $64.55$~GeV of the $4d$ theory without top for the lower
and larger $M_H^*$.
The four dimensional $\overline{\mathrm {MS}}$ gauge coupling $g^2(m_W)$ has the value 0.423
and 0.422, respectively, which is close to that of the standard model.

Our present result for the case of $M_H^*=70$~GeV is larger by a factor $3.3$
than the previous rough estimate \cite{wirNP97}. The latter was obtained as a
result of our equal weight histogram method, finally based on a global infinite
volume extrapolation of the mixed phase thermodynamical weight for lattices of
various aspect ratios.

In the present fits we obtain effective exponents $\gamma=0.25(11)$ and
$0.095(45)$ of $L$, respectively, which become smaller with decreasing strength
of the transition.  Fixing the exponent to $\gamma = 1$ as suggested by
eq.~(\ref{eq:three}) we would be able to present only a local fit to the few
highest transverse areas $L^2$. The interface tensions would be estimated by
this fit as follows: $\alpha_3/g_3^4=0.023$ for $M_H^*=57$~GeV and $0.016$ for
$M_H^*=70$~GeV.  In the first case this would be still acceptable comparing with
the result of the Helsinki group \cite{KajantieNP96} for that Higgs mass.  But
the interface tension evaluated at $M_H^*=70$~GeV in this way does not follow
the general trend of the interface tensions from $d=3$ simulations which have
been collected in Ref.~\cite{rummustlouis}.

\begin{figure}[tb]
  \centering
  \epsfig{file=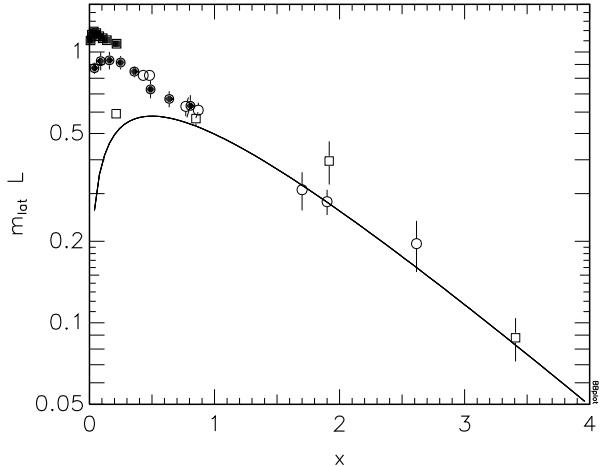,width=8.0cm,angle=0}
  \caption{\sl Scaling law for the mass gap
    followed by data at 
    $m_H=35$~GeV (open squares) [9],
    $m_H=49$~GeV (open circles) [8],
    $M_H^*=57$~GeV (full circles) and
    $M_H^*=70$~GeV (full squares) [this work]}
  \label{fig:four}
\end{figure}

In view of the arguments above, we consider the fit with the free ansatz
(\ref{eq:four}) more serious than the results of the fit confined to $\gamma=1$.
We recall that also in Ref.~\cite{bunk94} the latter fit has been successful
only at large values of the scaling variable $x={\alpha L^2 a^2 }/{T_c}$.  The
one--loop scaling law (\ref{eq:scaling}) was meant to hold independent of the
particular system, irrespective of the actual value of the interface tension.
With the values of $\alpha$ obtained now and in Refs.~\cite{bunk94,csikorpl95}
the underlying mass gaps for various Higgs masses and transverse lattice sizes
seem to follow another universal law if expressed through the scaling variable
$x$
\begin{equation}
  m_{\mathrm lat}~L = C^{\prime} ~\exp (- x)  
  \label{eq:scal1}
\end{equation}
instead of (\ref{eq:scal1}), except for the smallest transverse extensions $L$
in each case.  This can be seen in Fig.~\ref{fig:four}.  In this figure we also
show the asymptotic behaviour (\ref{eq:scaling}) with $C=1.352$ as predicted by
\cite{muenster,bunk}.

There is one warning in the data concerning the applicability of the analytical
formulae used throughout the literature in order to extract the interface
tension from lattice data of the tunneling correlation length.  As mentioned
above it is implicitly assumed that the correlation lengths of the two phases in
equilibrium are small compared to the tunneling correlation length. From this
point of view the present data on $m_{\mathrm lat}$ for $M_H^*=70$~GeV (where we have
measured Higgs masses separately for both phases at the transition temperature)
indicate that transverse sizes $L\leq 10$ would have to be excluded from the
analysis.

The other concern is caused by the fact that the {\it transverse} correlation
length $\xi_{\mathrm \perp}$ does not decouple from the transverse lattice
extension at small $L$ and keeps growing for all transverse sizes considered.
For instance, at $M_H^*=70$~GeV, the transverse correlation length amounts from
$75$ to $83$ per cent of $L$ on our lattices with transverse sizes $L \geq 10$
which are, on the other hand, the only acceptable ones in view of the criticism
discussed before.  In the case of $M_H^*=57$~GeV the transverse correlation
length is even larger compared with the transverse size $L$ ($100$ to $130$ per
cent) on lattices with $L \geq 10$.  It should be mentioned that the ratio of
the transverse correlation length to the transverse size is consistent with the
ratio of the bulk correlation length to the system size for cubic symmetries on
top of the phase transition (measured without separating the Monte Carlo
sample into pure phase configurations).

\section{Overview and Conclusions}

To compare our $3d$ results for the interface tension with those
of $4d$ measurements we follow the procedure outlined in \cite{laine}.
The measurements in the $4d$ approach have been performed at a different gauge
coupling. The  measured renormalised gauge couplings do not seem to change
significantly with the Higgs mass in the so far reported region 
from 18 to 49 GeV \cite{csikor,desy}
and vary from 0.56 to 0.59.
For simplicity (and due to missing calculation) it is assumed 
as in \cite{laine} that the measured renormalised  
gauge coupling roughly corresponds to the 
$\overline{\mathrm{MS}}$ running coupling. It is then
different from the value $g^2(m_W)=0.42$ corresponding to our calculations
described in the last section.

Using the prescription to relate $3d$ and $4d$ parameters \cite{wirNP97}
we calculate  Higgs masses, critical temperatures and  
interface tensions for increasing $4d$ gauge couplings
keeping the respective $\lambda_3/g_3^2$ fixed. 
The numbers are collected in table~\ref{tab:st_of_g}.
\begin{table}[htbp]
  \begin{center}
    \leavevmode
    \begin{tabular}{|c|c|c|c|c|}
      \hline
       & & & & \\[-2ex]
    $M_H^*/$GeV &       $g^2(m_W)$   & $m_H/$GeV&  $T_c/$GeV  &   $\alpha/T_c^3$   \\[0.5ex]
      \hline
       & & & & \\[-2ex]
      57        &     0.423        & 52.35      & 127.2    &   0.00324(081)     \\
                &     0.560        & 50.86      & 108.2    &   0.00539(134)     \\
                &     0.570        & 50.75      & 107.1    &   0.00556(138)     \\
                &     0.580        & 50.64      & 106.0    &   0.00574(143)     \\
                &     0.590        & 50.54      & 104.9    &   0.00592(147)     \\[0.5ex]
      \hline
       & & & & \\[-2ex]
      70        &     0.422        & 64.55      & 150.1    &   0.00070(26)    \\
                &     0.560        & 62.99      & 128.1    &   0.00115(43)    \\
                &     0.570        & 62.88      & 126.8    &   0.00119(45)    \\
                &     0.580        & 62.76      & 125.5    &   0.00122(46)    \\
                &     0.590        & 62.65      & 124.3    &   0.00126(47)    \\[0.5ex]
      \hline
    \end{tabular}
    \caption{\sl Higgs masses, critical temperatures and interface tensions for various $4d$ 
             running gauge couplings}
    \label{tab:st_of_g}
  \end{center}
\end{table}
Note that the Higgs mass is slightly moving to lower values while 
the ratio $\alpha/T_c^3$ becomes
much bigger, largely due to the critical temperature getting smaller.

\begin{figure}[!htb]
  \centering
  \epsfig{file=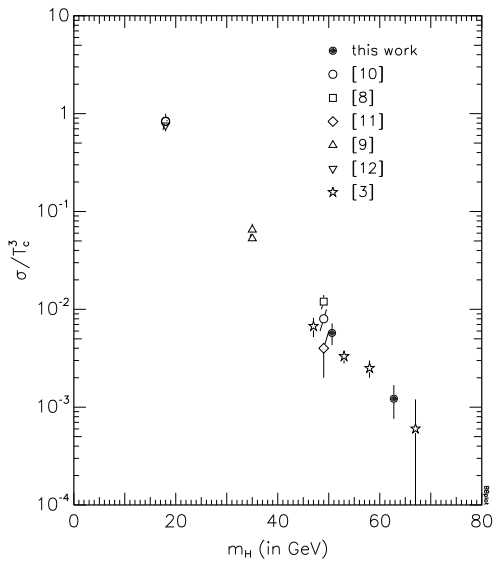,width=6.5cm,angle=0}
  \caption
  {\sl Collected results on the interface tension as function of the Higgs mass}
  \label{fig:five}
\end{figure}

Having the caveats of the last section in mind we compare 
now in Fig.~\ref{fig:five} our interface tensions $\alpha$ 
with those measured
by different methods in the $4d$ theory at various Higgs masses as 
function of the physical Higgs mass $m_H$.
In order to do this we read for our  $3d$ data (full points) 
the corresponding Higgs mass (without fermion contributions) 
from table~\ref{tab:st_of_g}, 
for example at $g^2(m_W)=0.58$. This value has been deduced 
as mentioned above from the measured renormalised  
gauge coupling.

In the $4d$ data the
Higgs masses are either measured \cite{csikor,bunk94,emias94,csikorpl95,hein}
(no errors are taken into account in the horizontal scale) or estimated
\cite{aoki}.  The agreement between the various approaches in $3d$ and
$4d$ is remarkable if 
the $3d$ and $4d$ parameters are correctly mapped onto each other. This nicely
demonstrates the validity of dimensional reduction in the considered
Higgs mass range.

Our new data points are the following ones, expressed in terms of the
$3$--dimensional and the $4$--dimensional interface tensions.  We find at
$\lambda_3/g_3^2=0.0644457$, $g^2(m_W)=0.423$ and $\beta_G=12$ ($M_H^*=57$~GeV)
\begin{equation}
  \frac{\alpha_3}{g_3^4}=0.0224(56),\ \ \ \frac{\alpha}{T_c^3}=3.24(80) \times
  10^{-3}
  \label{eq:r1}
\end{equation}
and at $\lambda_3/g_3^2=0.095703$, $g^2(m_W)=0.422$ and $\beta_G=12$ ($M_H^*=70$~GeV)
\begin{equation}
  \frac{\alpha_3}{g_3^4}=0.0049(18),\ \ \ \frac{\alpha}{T_c^3}=70(26) \times
  10^{-5}    .
  \label{eq:r2}
\end{equation}

We observe an approximate scaling law expressing the energy gap for all Higgs
masses in terms of the dimensionless variable $x$ (see eq.~(\ref{eq:three1}))
without the prefactor $\sqrt x$ which had been suggested by one--loop
perturbation theory for interfaces in the case of binary systems.

We emphasise that the tunneling correlation length method works well even near
to the critical Higgs mass where other methods relying on discrimination of
histogram peaks and minima are already difficult to apply.

\section*{Acknowledgements}

E.M.~I. is supported by the DFG under grant Mu932/3-4.  We thank the council
of HLRZ J\"ulich for providing CRAY-T90 resources.


\begin{thebibliography}{99}

\bibitem{kajantieprl96} 
  K.~Kajantie, M.~Laine, K.~Rummukainen and M.~Shaposhnikov,
  Phys. Rev. Lett. {\bf 77} (1996) 2887
 
\bibitem{karsch96}
  F.~Karsch, T.~ Neuhaus, A.~Patk\'os and J.~Rank, 
  Nucl. Phys. {\bf B}(Proc. Suppl.) {\bf 53} (1997) 623

\bibitem{aoki}  
  Y.~Aoki,
  Nucl. Phys. {\bf B}(Proc. Suppl.) {\bf 53} (1997) 609 
  and hep-lat/9612023
  
\bibitem{preparation}
  M.~G\"urtler, E.-M.~Ilgenfritz and A.~Schiller, in preparation

\bibitem{wirNP97} 
  M.~G\"urtler, E.-M.~Ilgenfritz, J.~Kripfganz, H.~Perlt and A.~Schiller,
  Nucl. Phys. {\bf B483} (1997) 383
  
\bibitem{privman} 
  V.~Privman and  M.E.~Fisher, J. Stat. Phys. {\bf 33} (1983) 385

\bibitem{muenster}  
  G.~M\"unster, Nucl. Phys. {\bf B340} (1990) 559  

\bibitem{bunk94} 
  B.~Bunk,
  Nucl. Phys. {\bf B}(Proc. Suppl.) {\bf 42} (1995) 566

\bibitem{csikorpl95}
  F.~Csikor, Z.~Fodor, J.~Hein, J.~Heitger,
  Phys. Lett. {\bf B357} (1995) 156

\bibitem{csikor}
  F.~Csikor, Z.~Fodor, J.~Hein, K.~Jansen, A.~Jaster and I.~Montvay, 
  Phys. Lett {\bf B334} (1994) 405;
  Z.~Fodor, J.~Hein, K.~Jansen, A.~Jaster and I.~Montvay,
  Nucl. Phys. {\bf B439} (1995) 147

\bibitem{emias94} 
  E.-M.~Ilgenfritz and A.~Schiller, 
  Nucl. Phys. {\bf B}(Proc. Suppl.) {\bf 42} (1995) 578

\bibitem{hein}
  J.~Hein, J.~Heitger, Phys. Lett. {\bf B385} (1996) 242

\bibitem{KajantieNP96} 
  K.~Kajantie, M.~Laine, K.~Rummukainen and M.~Shaposhnikov,
  Nucl. Phys. {\bf B466} (1996) 189

\bibitem{iwasaki} 
  Y.~Iwasaki, K.~Kanaya, L.~K\"arkk\"ainen, K.~Rummukainen and T.~Yoshi\'e,
  Phys. Rev. {\bf D49} (1994) 3540

\bibitem{brezin}
 E.~Br\`ezin and J.~Zinn-Justin, Nucl. Phys. {\bf B257} (1985) 867
  
\bibitem{private_muenster}   
 G.~M\"unster, private communication  

\bibitem{caselle} 
  M.~Caselle, R.~Fiore, F.~Gliozzi, M.~Hasenbusch, 
  K.~Pinn and S.~Vinti, Nucl. Phys. {\bf B432} (1994) 590

\bibitem{rummustlouis} 
  K.~Rummukainen,
  Nucl. Phys. {\bf B}(Proc. Suppl.) {\bf 53} (1997) 30

\bibitem{bunk}  
  B.~Bunk,
  Int. J. Mod. Phys. {\bf C3} (1992) 889

\bibitem{laine}  
  M.~Laine,
  Phys. Lett. {\bf B385} (1996) 249

\bibitem{desy}
  F.~Csikor, Z.~Fodor, J.~Hein,  A.~Jaster and I.~Montvay,
  Nucl. Phys. {\bf B474} (1996) 421

\end{thebibliography}
\end{document}